\documentclass[aps,preprint,onecolumn,superscriptaddress]{revtex4-1}
\usepackage{amsmath}
\usepackage{graphicx}
\usepackage{float}
\DeclareGraphicsExtensions{.pdf,.eps,.png,.jpg,.mps}
\usepackage{epstopdf}
\usepackage{threeparttable}
\usepackage{xcolor}
\usepackage{soul}
\usepackage{caption}
\begin{document}
\title{Confinement of long-lived interlayer excitons in WS$_2$/WSe$_2$ heterostructures}

\author{Alejandro R.-P. Montblanch}\thanks{These two authors contributed equally.}
\affiliation{\footnotesize{\hbox{Cavendish Laboratory, University of Cambridge, 19 J. J. Thomson Ave., Cambridge, CB3 0HE, UK}}}
\author{Dhiren M. Kara}\thanks{These two authors contributed equally.}
\affiliation{\footnotesize{\hbox{Cavendish Laboratory, University of Cambridge, 19 J. J. Thomson Ave., Cambridge, CB3 0HE, UK}}}
\author{Ioannis Paradisanos}
\affiliation{\footnotesize{\hbox{Cambridge Graphene Centre, University of Cambridge, 9 J. J. Thomson Ave., Cambridge, CB3 0FA, UK}}}
\author{Carola M. Purser}
\affiliation{\footnotesize{\hbox{Cavendish Laboratory, University of Cambridge, 19 J. J. Thomson Ave., Cambridge, CB3 0HE, UK}}}
\affiliation{\footnotesize{\hbox{Cambridge Graphene Centre, University of Cambridge, 9 J. J. Thomson Ave., Cambridge, CB3 0FA, UK}}}
\author{Matthew S. G. Feuer}
\affiliation{\footnotesize{\hbox{Cavendish Laboratory, University of Cambridge, 19 J. J. Thomson Ave., Cambridge, CB3 0HE, UK}}}
\author{Evgeny M. Alexeev}
\affiliation{\footnotesize{\hbox{Cambridge Graphene Centre, University of Cambridge, 9 J. J. Thomson Ave., Cambridge, CB3 0FA, UK}}}
\affiliation{\footnotesize{\hbox{Cavendish Laboratory, University of Cambridge, 19 J. J. Thomson Ave., Cambridge, CB3 0HE, UK}}}
\author{Lucio Stefan}
\affiliation{\footnotesize{\hbox{Cavendish Laboratory, University of Cambridge, 19 J. J. Thomson Ave., Cambridge, CB3 0HE, UK}}}
\affiliation{\footnotesize{\hbox{The Faraday Institution, Quad One, Becquerel Avenue, Harwell Campus, Didcot, OX11 0RA, UK}}}
\author{Ying Qin}
\affiliation{\footnotesize{\hbox{School for Engineering of Matter, Transport and Energy, Arizona State University, Tempe, AZ, 85287, USA}}}
\author{Mark Blei}
\affiliation{\footnotesize{\hbox{School for Engineering of Matter, Transport and Energy, Arizona State University, Tempe, AZ, 85287, USA}}}
\author{Gang Wang}
\affiliation{\footnotesize{\hbox{Cambridge Graphene Centre, University of Cambridge, 9 J. J. Thomson Ave., Cambridge, CB3 0FA, UK}}}
\author{Alisson R. Cadore}
\affiliation{\footnotesize{\hbox{Cambridge Graphene Centre, University of Cambridge, 9 J. J. Thomson Ave., Cambridge, CB3 0FA, UK}}}
\author{Pawel Latawiec}
\affiliation{\footnotesize{\hbox{School of Engineering and Applied Science, Harvard University, Cambridge, MA, 02138, USA}}}
\author{Marko Lon\v{c}ar}
\affiliation{\footnotesize{\hbox{School of Engineering and Applied Science, Harvard University, Cambridge, MA, 02138, USA}}}
\author{Sefaattin Tongay}
\affiliation{\footnotesize{\hbox{School for Engineering of Matter, Transport and Energy, Arizona State University, Tempe, AZ, 85287, USA}}}
\author{Andrea C. Ferrari}\email{acf26@cam.ac.uk}
\affiliation{\footnotesize{\hbox{Cambridge Graphene Centre, University of Cambridge, 9 J. J. Thomson Ave., Cambridge, CB3 0FA, UK}}}
\author{Mete Atat\"ure}\email{ma424@cam.ac.uk}
\affiliation{\footnotesize{\hbox{Cavendish Laboratory, University of Cambridge, 19 J. J. Thomson Ave., Cambridge, CB3 0HE, UK}}}

\begin{abstract}
Interlayer excitons in layered materials constitute a novel platform to study many-body phenomena arising from long-range interactions between quantum particles. The ability to localise individual interlayer excitons in potential energy traps is a key step towards simulating Hubbard physics in artificial lattices. Here, we demonstrate spatial localisation of long-lived interlayer excitons in a strongly confining trap array using a WS$_{2}$/WSe$_{2}$ heterostructure on a nano-patterned substrate. We detect long-lived interlayer excitons with lifetime approaching 0.2 ms and show that their confinement results in a reduced lifetime in the microsecond range and stronger emission rate with sustained optical selection rules. The combination of a permanent dipole moment, spatial confinement and long lifetime places interlayer excitons in a regime that satisfies one of the requirements for observing long-range dynamics in an optically resolvable trap lattice.
\end{abstract}


\maketitle

Since the demonstration of Bose-Einstein condensation\cite{Anderson1995,Davis1995}, ultracold atoms have played a central role in the study and realization of macroscopic-scale quantum phenomena, including Mott transition\cite{Greiner2002}, superfluidity\cite{Zweierlein2005} and many-body localisation\cite{Schreiber2015}. In parallel, exciton-polaritons have become an equivalent platform in semiconductors\cite{Deng2002,Kasprzak2006,Balili2007,Plumhof2013,Byrnes2014,Wertz2010,Amo2009}. Recently, dipolar particles have gained attention as they introduce long-range anisotropic interactions to the exploration of new states of quantum matter\cite{Baranov2012,Trefzger2011,Lahaye2009}. Atomic realizations of dipolar ensembles include Rydberg atoms\cite{Schauss2012,Schauss2015,Saffman2010}, ultracold polar molecules\cite{Truppe2017,Bohn2017,Carr2009} and high magnetic-moment atoms\cite{Griesmaier2005,Lu2011,Aikawa2012,Baier2016,Paz2013}. For semiconductors, long-range interactions can be achieved via spatially indirect excitons, where electrons and holes have finite separation\cite{Chen1987,Golub1988} yielding a permanent electric dipole moment\cite{Voros2005,Hubert2019,Hubert2019b}. Most progress in realizing such states has been in AlGaAs/GaAs double quantum wells\cite{Butov2002,High2012,Shilo2013,Stern2014,Hubert2019}, while heterostructures of transition metal dichalcogenide (TMD) monolayers\cite{Rivera2015} have recently emerged as a promising alternative\cite{Miller2017,Rivera2018,Forg2019,Jauregui2019}. TMD monolayers are semiconductors offering optical access to orbital, spin and valley degrees of freedom\cite{Wang2018}, and the difference in band energies of two different TMDs is exploited to create type-II interlayer excitons\cite{Kang2013,Gong2013} with static electric dipole. Moreover, their optical transition strength, energy, and selection rules can be engineered by controlling the layer separation\cite{Fang2014} and the relative stacking angle of the monolayers\cite{Heo2015,Nayak2017,Alexeev2019,Seyler2019,Jin2019,Tran2019}.

The potential of TMD interlayer excitons for many-body physics is evidenced by recent reports of exciton condensation\cite{Wang2019} and the creation of moiré lattices for exploring Hubbard\cite{Tang2019} and Mott\cite{Shimazaki2019,Regan2019} physics. While a moiré lattice achieved through angled stacking of TMD monolayers is an exciting route, the typical lattice constant is restricted to a few nanometres, well below the optical diffraction limit ($\sim$0.5 $\mu$m), preventing single-site optical access. Trapping interlayer excitons in an independent potential energy landscape offers a solid-state analogue of atoms in single-site addressable optical lattices\cite{Bakr2009,Sherson2010}, and first indications of confined interlayer excitons were reported recently\cite{Kremser2019,Li2019}. An elementary requirement for long-range interactions to manifest in such systems is that the interlayer exciton decay rate is smaller than the dipole-dipole coupling rate in neighbouring lattice sites. For TMD interlayer excitons, shown to have an electric dipole of $\sim$0.6 e$\cdot$nm ($\sim$29 Debye)\cite{Jauregui2019}, the dipole-dipole coupling rate at a distance of $\sim$0.5 $\mu$m is 1 MHz\cite{Blackmore2019}. This means that a trapped interlayer exciton lifetime of at least one microsecond needs to be achieved.

Figure 1a is an illustration of our device, fabricated by exfoliating WSe$_{2}$ and WS$_{2}$ monolayers from bulk crystals and transferring these onto a SiO$_{2}$/Si substrate patterned with an array of nanopillars 220-250 nm tall and 4 $\mu$m apart (see Methods and Section 1 of the Supplementary Information, SI). We choose an interlayer stacking angle of $\sim$50$^{º}$, away from high-symmetry angles of 0$^{º}$ or 60$^{º}$, in order to introduce an interlayer momentum mismatch to reduce the exciton recombination rate. Five distinct locations are labelled in Fig. 1a as L1-L5. These locations are marked by circles in the optical image of our device shown in Fig. 1b, where the WSe$_2$ and WS$_2$ monolayers are outlined in red and green, respectively, yielding a large region ($>$160 $\mu$m$^{2}$) of the WS$_{2}$/WSe$_{2}$ heterobilayer.

Figure 1c shows a photoluminescence (PL) map taken with continuous-wave (CW) 2.33-eV (532-nm) laser excitation at a temperature T=4 K. This energy is greater than both the WSe$_2$ and WS$_2$ optical bandgaps of 1.73 eV\cite{Jones2013} and 2.08 eV\cite{Mitioglu2013}, respectively. Consequently, PL emission is observed from both monolayer and heterobilayer regions. The spectral emission from the flat monolayer regions (L1 and L2 in Fig. 1b) is shown in Figs. 1d (i) and (ii), respectively. The PL emission between 1.65 and 2.0 eV in panel (i) (filled in green) is the intralayer exciton recombination in monolayer WS$_2$\cite{Mitioglu2013,Palacios2017}, and that between 1.6 and 1.75 eV in panel (ii) (filled in red) is the intralayer exciton recombination in monolayer WSe$_2$\cite{Jones2013,Barbone2018}. Figure 1d panel (iii) shows the spectrum from the flat part of the heterobilayer, L3 in Fig. 1b. A lower-energy feature around 1.4 eV (filled in yellow) is consistent with interlayer exciton emission from stacked WS$_{2}$/WSe$_{2}$ heterobilayers reported previously\cite{Jin2019,Jin2019b}. Figure 1d (iv) is the PL spectrum from a monolayer WSe$_{2}$ on a nanopillar, presented as reference, displaying a $\sim$50-fold brighter emission with respect to flat monolayer WSe$_{2}$ and a sub-meV full-width at half-maximum spectral peak. Given that monolayer WSe$_2$ on such nanopillars results in the strong quantum confinement of intralayer excitons\cite{Palacios2017,Branny2017}, we look for similar signatures in the interlayer emission spectra. Figure 1d (v) is the PL spectrum of the heterobilayer on a nanopillar. We observe a 20-fold brighter interlayer emission in the low optical excitation regime (below 1 $\mu$W) and sharp spectral features ($\sim$1 meV full-width at half-maximum), in contrast to the weak and spectrally broad emission from the flat heterobilayer.

To isolate interlayer excitations from any effects coming from the monolayers, such as free charge carriers or intralayer excitons, we use an optical excitation energy of 1.50 eV for all other measurements. This energy is at the onset of the interlayer exciton PL spectrum, Fig. 1d (iii), and below the optical bandgap energies of WSe$_2$ and WS$_2$ monolayers. Figure 1e displays the PL intensity map of our device under 1.50-eV excitation. The emission is only observed from the heterobilayer region with both monolayer regions remaining dark, demonstrating that only interlayer excitons are generated, as anticipated.

Figure 2a shows the interlayer exciton PL spectra from the flat heterobilayer for 0.05-$\mu$W (black curve), 1-$\mu$W (red curve) and 100-$\mu$W (blue curve) laser excitation power, \textit{P}, at T=4 K. The spectrally integrated PL intensity of interlayer exciton with respect to \textit{P} is shown in Fig. 2b. While the interlayer exciton emission starts with a linear dependence on \textit{P}, it converges to a \textit{P}$^{0.3}$ scaling for \textit{P}$>$0.3 $\mu$W. Sublinear behaviour was also reported for MoSe$_2$/WSe$_2$ heterobilayers\cite{Rivera2015}. This can be caused by density-dependent mechanisms such as exciton-exciton annihilation, dipolar repulsion and phase-space filling.

Figure 2c presents PL spectra from the heterobilayer on a nanopillar for \textit{P}=0.05 $\mu$W (black curve), 1 $\mu$W (red curve) and 100 $\mu$W (blue curve). With increasing power the three peaks highlighted in green, grey and blue saturate. The integrated PL intensity as a function of \textit{P} from these peaks are plotted in Fig. 2d, with data (squares) colour-coded for the highlighted peaks of Fig. 2c. Each shows a linear \textit{P}-dependence, followed by saturation. This behaviour is characteristic of a quantum-confined system\cite{Kurtsiefer2000,He2015}, and contrasts that of untrapped interlayer excitons in the flat heterobilayer (Figs. 2a and b). Eight nanopillar locations display this saturating behaviour (Fig. S3a in SI), demonstrating that interlayer-exciton trapping is reproduced across the device.

Figures 3a and 3b present trapped interlayer exciton PL spectra above a nanopillar (L5 of Fig. 1b) at T=4 K, for out-of-plane magnetic field \textit{B} from 0 to 9 T. In panel a (panel b) the excitation is linearly polarised and the collection is right-hand $\sigma^{+}$ (left-hand $\sigma^{-}$) circularly polarised. The red-shifting Zeeman state from each emission line is only seen for $\sigma^{+}$ detection, while the blue-shifting Zeeman state is only seen for $\sigma^{-}$ detection. This demonstrates that the two magnetic configurations of trapped interlayer excitons have sustained optical selection rules\cite{Aivazian2015}.

The distribution of Zeeman splitting with respect to \textit{B} across different nanopillars is given in Fig. 3c as a blue-shaded region, where the data corresponding to the smallest and largest measured g-factor (11.9 and 15.4) are plotted in blue and red circles, respectively.  The similarity in g-factors, mean value 13.2$\pm$1.1, across the nanopillar traps and for different PL peaks (Fig. S3b in SI) suggests that the trapped excitons have the same microscopic origin. The variation in g-factors is also similar to that observed in confined intralayer excitons in WSe$_{2}$ monolayers\cite{Srivastava2015,He2015,Koperski2015,Chakraborty2015}. This contrasts the order of magnitude smaller distribution in g-factor observed for trapped interlayer excitons in homogeneous moiré trapping potentials\cite{Seyler2019}.

Figure 3d presents the PL spectra under $\sigma^{+}$ excitation at \textit{B}=0 T for $\sigma^{+}$ (black) and $\sigma^{-}$ (red) circularly polarised collection. The overlap of the two spectra demonstrates that excitation polarisation is not maintained for trapped interlayer excitons. One dominant mechanism for this loss of polarisation is the exchange interaction between bound electron and holes\cite{Glazov2014}. For trapped interlayer excitons the exchange interaction rate and the exciton decay rate should be reduced proportionally with the increased electron-hole separation, but the exciton decay rate is additionally suppressed when the momentum-space overlap is reduced, as expected for heterobilayers with a stacking angle not matching 0$^{º}$ or 60$^{º}$\cite{Nayak2017}.

Figure 4 presents lifetime measurements on interlayer excitons under pulsed ($\sim$3-ps pulse duration) laser excitation at 1.50 eV at T=4 K (see Methods). Figure 4a shows an example emission intensity histogram of untrapped interlayer excitons (grey bars) as a function of time after excitation, and an exponential fit (solid red curve) reveals a decay time of 180.6$\pm$3.5 $\mu$s. Across the flat heterobilayer region, the average lifetime is 175$\pm$5 $\mu$s. Since only interlayer excitons are created under 1.50 eV excitation, spuriously prolonged lifetimes resulting from the delayed capture of free carriers\cite{Shamirzaev2004} or higher energy intralayer dark excitons\cite{Jiang2018} are avoided. The spatial and momentum separation of electrons and holes of interlayer excitons, and the use of high-quality monolayers (see Methods), are likely responsible for these values. That said, this lifetime shows a strong excitation dependence: increasing the excitation power reduces the exciton decay time (Fig. S4 in SI), as expected from density-dependent interactions and loss channels, even in the regime where PL intensity depends linearly on the excitation power (\textit{P}$<$0.3 $\mu$W in Fig. 2b).

Figure 4b is a lifetime measurement of the trapped interlayer exciton spectral line at 1.39 eV from Fig. 3d (L5 in Fig. 1b). We fit a biexponential function of the form A$_s$e$^{-t/\tau_{s}}$+A$_l$e$^{-t/\tau_{l}}$, where $\tau_{l}$ ($\tau_{s}$) is the long (short) lifetime and $A_{l}$ ($A_{s}$) is the amplitude of the exponential with long (short) lifetime. This yields $\tau_s=$59.4$\pm$0.9 ns and $\tau_{l}=$389.1$\pm$1.2 ns. This biexponential behaviour, albeit with varying lifetimes, is observed throughout the entire spectral range for trapped interlayer excitons (Fig. S5 in SI). Figure 4c is a summary of the spectrally integrated lifetimes from the eight nanopillar traps. The extracted $\tau_s$ and $\tau_l$ values are shown in Fig. 4c in black and grey bars, and lie in the $\sim$10-175 ns and $\sim$0.4-4 $\mu$s ranges, respectively. The reduction in the interlayer exciton lifetime after trapping accompanies an enhancement in PL brightness, suggesting a modified oscillator strength under localisation. This likely arises from the relaxation of stacking-angle-induced momentum mismatch on nanopillars, that otherwise inhibits the recombination of untrapped interlayer excitons\cite{Feierabend2017}. One possible origin of the observed biexponential decay is the presence of two excited states: an optically active state and an energetically similar shelving state (inset diagram in Fig. 4b). A three-level model combined with temperature-dependent measurements (Section 5 in SI) yields average radiative and non-radiative recombination rates of trapped interlayer excitons across the nanopillar sites $1/\tau_{s}\approx\Gamma_{r}\approx$13 MHz and $1/\tau_{l}\approx\Gamma_{nr}\approx$500 kHz, respectively. This reveals the considerably smaller average non-radiative decay rate and indicates that the long-lived shelving state decays primarily through radiative emission. Temperature-dependent measurements demonstrate that thermal excitations also couple bright and shelving states as evidenced in Fig. S6d of SI.

The spatial trapping we show here is a first step towards building arrays of long-lived and interacting interlayer excitons. Our trapped interlayer exciton lifetime, $\sim$4 $\mu$s, is sufficient to observe dipole-mediated exciton interactions for an optically resolvable lattice spacing of $\sim$0.5 $\mu$m. The immediate next step is to develop a full account of the photophysics of these trapped interlayer excitons, including the influence of stacking angle, interlayer spacing, electrostatic gating of the heterobilayer, and resonant excitation. In addition, other trapping geometries, such as ridges and rings, can offer the opportunity to probe many-body phenomena in one dimension.

\section{Acknowledgements}

We acknowledge funding from the EU Quantum Technology (2D-SIPC) and Graphene Flagships; ERC grants PEGASOS, Hetero2D and GSYNCOR; EPSRC Grants EP/K01711X/1, EP/K017144/1, EP/N010345/1 and EP/L016057/1; and the Faraday Institution FIRG001. D. M. K. acknowledges support of a Royal Society university research fellowship URF\textbackslash R1\textbackslash 180593. S.T. acknowledges support from DOE, NSF DMR 1552220, DMR 1904716, and NSF CMMI 1933214.

\section{Methods}

\subsection{Fabrication}

Van der Waals TMD bulk crystals were synthesised through flux zone growth technique\cite{Zhang2015}. Precursor powders were purchased from commercial vendors (Alfa Aesar) but additional electrolytic purification process was implemented to achieve 99.9999\% or higher purity. After the purification these powders were analysed using secondary ion mass spectroscopy to confirm absence of metal impurities. Compared to crystals grown by chemical vapour transport, flux grown ones were free of point defects and topological defects.

Flakes are then prepared by micromechanical cleavage\cite{Novoselov2005} on Nitto Denko tape, then exfoliated again for transfer to a polydimethylsiloxane (PDMS) stamp placed on a transparent glass slide, allowing bidirectional inspection of the flakes under the optical microscope. Optical contrast is utilized to identify monolayers prior to transfer\cite{Casiraghi2007}. We get monolayer flakes with lateral dimensions $\sim$50 $\mu$m. Substrates with arrays of silica nanopillars, 100 nm in diameter and 220-250 nm high are prepared by direct-write lithography\cite{Palacios2017}. The substrates undergo wet cleaning (1-minute ultrasonication in acetone and isopropanol) and are subsequently exposed to an oxygen-assisted plasma at a power of 10 W for 60 s to remove impurities and contaminants from the surface. The WSe$_2$ monolayer is then stamped on the nanopillars with a micro-manipulator \cite{Palacios2017,Purdie2019}. After the first transfer, the second (WS$_2$) monolayer is deposited on top of the WSe$_{2}$ monolayer following the same stamping procedure. In both steps the PDMS stamp is removed after depositing the monolayer. Raman (Fig. S1a in SI), second-harmonic generation (Fig. S1b in SI) and atomic force microscopy (Fig. S2 in SI) measurements confirm the monolayer nature of the constituents, the twist angle and the interlayer spacing, respectively.

\subsection{Photoluminescence measurements}

All PL measurements are done at T=4 K in a 9-T closed-cycle cryostat (Attocube). Excitation and collection pass through a home-built confocal setup with the sample in reflection geometry. CW illumination from either a 2.33 eV laser (Ventus) or a Ti:Sapphire laser (Mira 900) at 1.58 or 1.50 eV is used. The PL signal is spectrally filtered and sent to a 150-line grating spectrometer (Princeton Instruments).

\subsection{Lifetime measurements}

To perform lifetime measurements, we excite the samples every few $\mu$s (up to $\sim$10 $\mu$s for the trapped interlayer excitons, and up to 1 ms for the interlayer exciton) with $\sim$3 ps pulses from a Ti:Sapphire laser (Mira 900) tuned to 785 nm (1.58 eV) or 825 nm (1.50 eV). An acousto-optic modulator (AOM) down-samples the 76 MHz laser repetition rate to the kHz-MHz range required to measure up to hundreds of $\mu$s lifetimes. A time-to-digital converter (QuTau) with 81 ps timing resolution collects start-stop histograms with "start" triggered by the AOM pulse-picking and "stop" triggered by an avalanche photodiode (APD) output from single-photon detection of interlayer exciton PL. The converter remains idle until a subsequent "start" signal arrives. We ensure that we are not susceptible to spurious artefacts from start-stop measurements by measuring lifetime at a subset of locations using photon time-tagging and calculating lifetimes in post-processing. Both methods result in consistent lifetimes for trapped interlayer excitons, but time-tagging is required for the measurement of the unbound interlayer exciton lifetimes, due to its count rate being comparable to or even lower than the APD dark counts, as well as being at least one order of magnitude smaller than the pulse rate sent to the sample.

\begin{figure*}
\centerline{\includegraphics[width=110mm]{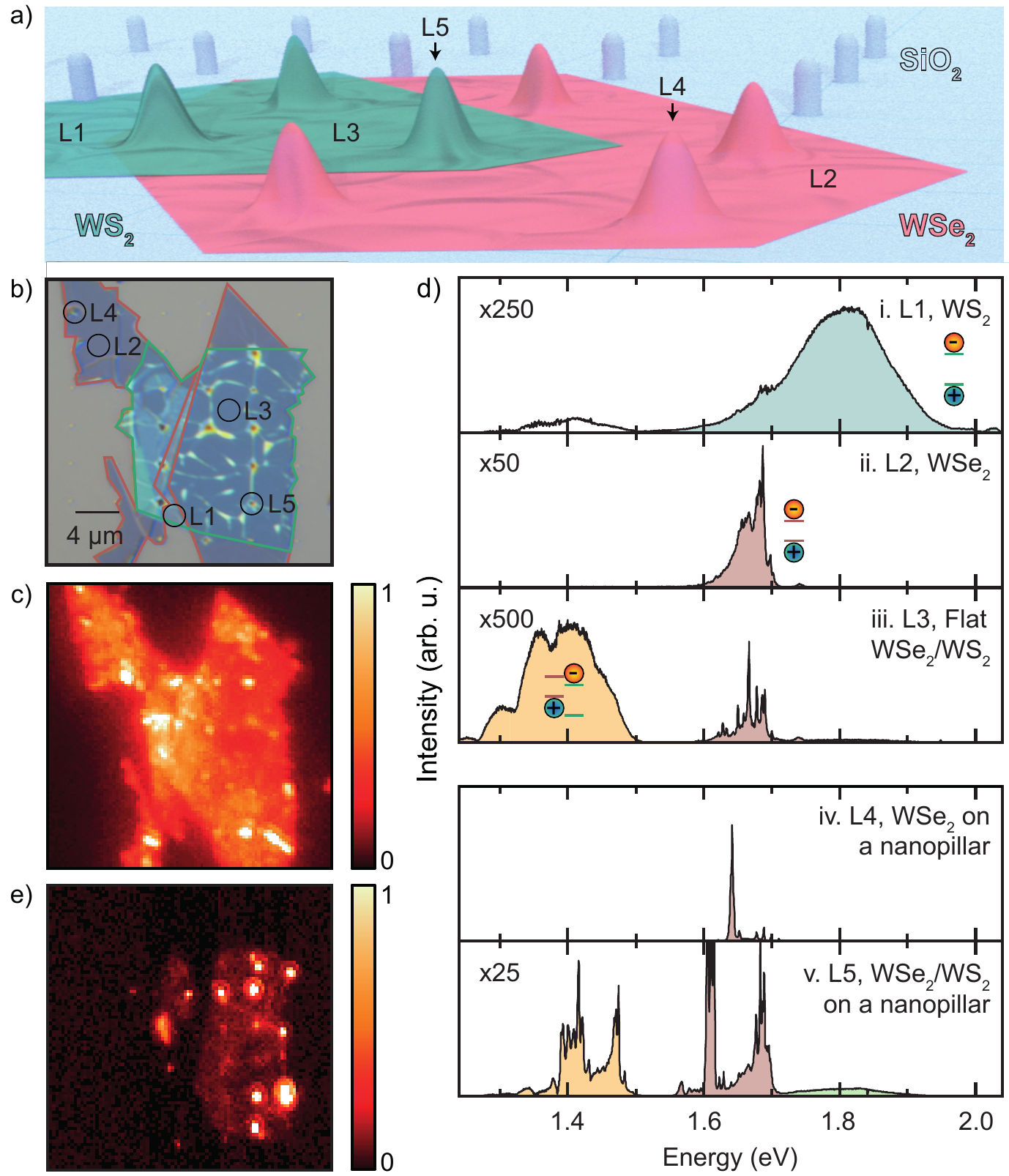}}
\caption{\textbf{Optical characterization of the WS$_{2}$/WSe$_{2}$ heterobilayer.}  \textbf{(a)} Illustration of our device. The SiO$_{2}$ substrate with nanopillars is in blue, the WSe$_{2}$ monolayer in red and WS$_{2}$ on top in green. Representative locations on the device are indicated: location L1 is the WS$_2$ monolayer, location L2 is the WSe$_2$ monolayer, location L3 is the flat WS$_{2}$/WSe$_{2}$ heterobilayer, location L4 is the WSe$_{2}$ monolayer on a nanopillar and location L5 is the WS$_{2}$/WSe$_{2}$ heterobilayer on a nanopillar. \textbf{(b)} Optical image of our device. WSe$_{2}$ and WS$_{2}$ monolayers are outlined in red and green, respectively. \textbf{(c)} Integrated PL intensity map of our device under 2.33-eV CW excitation at T=4 K. \textbf{(d)} Representative PL spectra under 2.33-eV CW excitation taken at the five locations highlighted in panel b. The colour coding indicates the origin of PL emission, where green (red) comes from intralayer excitons in WS$_{2}$ (WSe$_{2}$) monolayer, and yellow comes from interlayer excitons in the WS$_{2}$/WSe$_{2}$ heterobilayer. The multiplicative factors of each PL spectrum are all referenced to subpanel (iv), WSe$_{2}$ monolayer spectrum at L4. \textbf{(e)} PL intensity map of our device under 1.50-eV CW excitation at T=4 K, to which only the heterobilayer regions contribute.}
\label{fig:1}
\end{figure*}

\begin{figure*}
\centerline{\includegraphics[width=180mm]{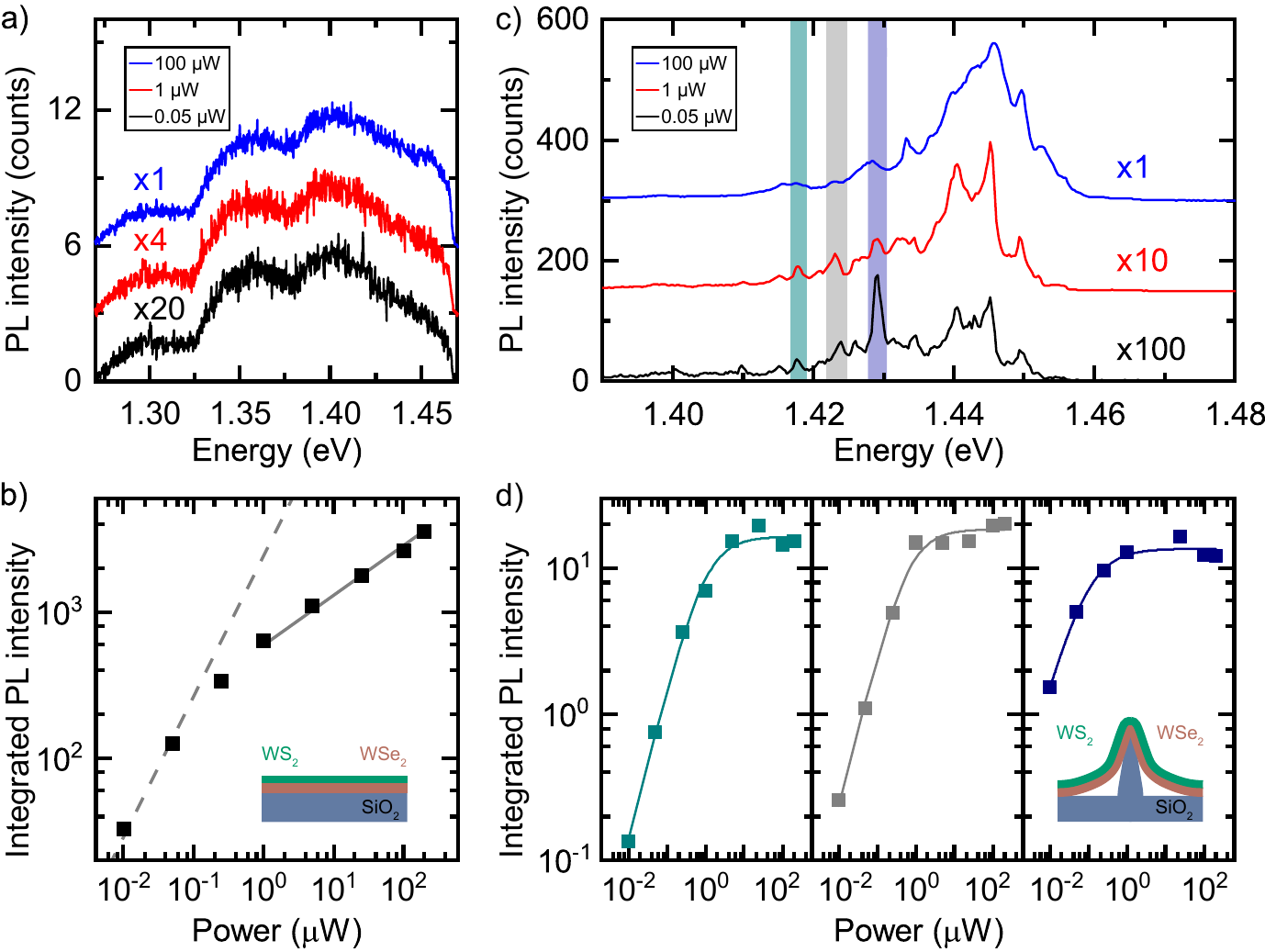}}
\caption{\textbf{Excitation-power dependence of interlayer exciton emission.} \textbf{(a)} PL spectra at T=4 K from the flat heterobilayer under 1.50-eV excitation at 0.05 $\mu$W (black curve), 1 $\mu$W (red curve) and 100 $\mu$W (blue curve). \textbf{(b)} Spectrally integrated PL intensity of interlayer exciton emission as a function of excitation power, \textit{P}. The dashed grey curve follows a linear dependence in \textit{P}, while the solid grey curve follows \textit{P}$^{0.3}$. \textbf{(c)} PL spectra from the heterobilayer on a nanopillar at 0.05-$\mu$W (black curve), 1-$\mu$W (red curve) and 100-$\mu$W (blue curve) excitation. \textbf{(d)} Spectrally integrated PL intensity of the three peaks shaded with green, grey and blue bands in panel c as a function of \textit{P}. Data (filled squares) are colour-coded to the spectral bands of panel c. Solid curves are fits to data using the saturation function $A\times P^{n}/(P_{\rm{sat}} + P^{n}$), from where we determine n=0.94$\pm$0.12, 1.06$\pm$0.08 and 1.04$\pm$0.06 and saturation powers of 0.1, 0.6, and 0.9 $\mu$W for the green, grey and blue data, respectively.}
\label{fig:2}
\end{figure*}

\begin{figure*}
\centerline{\includegraphics[width=160mm]{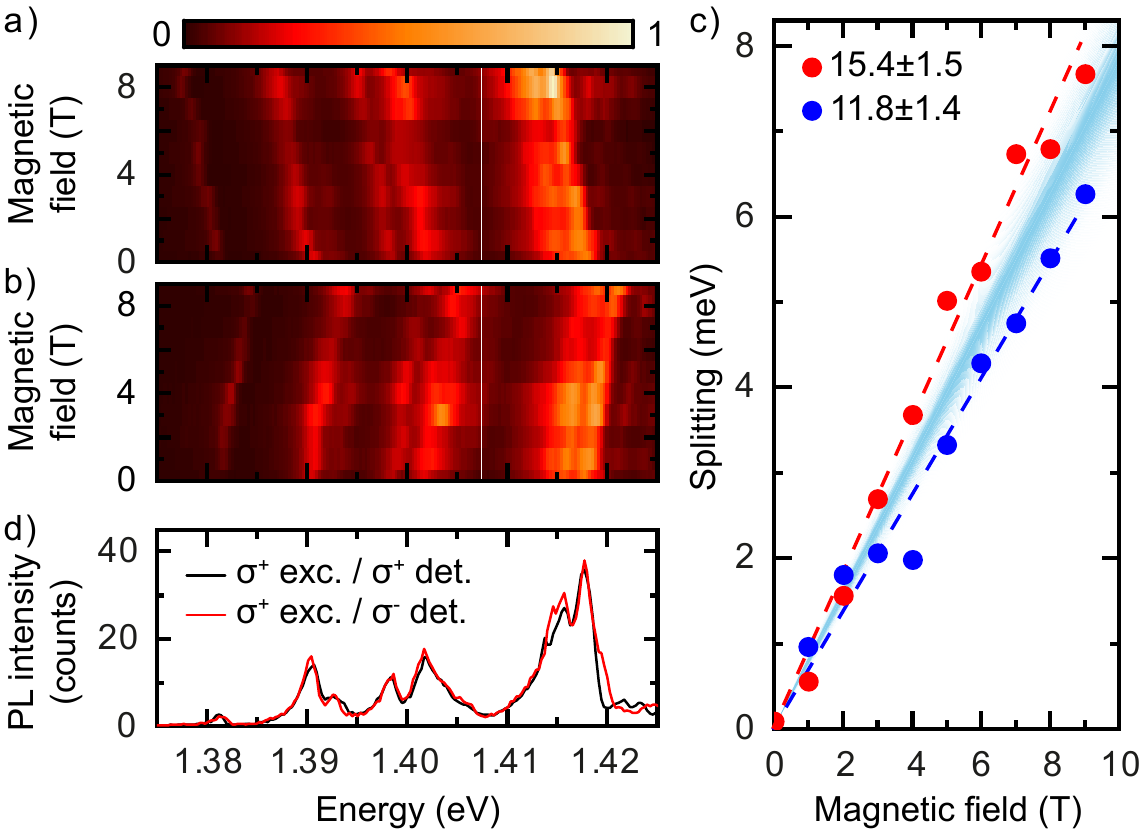}}
\caption{\textbf{Magnetic-field dependence of trapped interlayer excitons.} \textbf{(a)} PL spectrum at T=4 K as a function of magnetic field \textit{B} on a nanopillar under linearly polarised excitation and circularly polarised $\sigma^{+}$ detection. \textbf{(b)} Same measurement as panel a, but with $\sigma^{-}$ polarised detection. The intensity of the components in a increases with \textit{B}, and the blue-shifting components in b decrease with \textit{B}, consistent with exciton thermalisation of Zeeman states at T=4 K\cite{Barbone2018}. \textbf{(c)} Measured Zeeman splittings of trapped interlayer excitons as a function of \textit{B}. The red (blue) circles correspond to trapped interlayer excitons with the largest (smallest) g-factor of 15.4$\pm$1.5 (11.8$\pm$1.4) extracted from the linear fits (dashed curves). The mean g-factor value of all measured splittings is 13.2 with a standard deviation 1.1. When considering spin, valley and orbital contributions to the total magnetic moment, the measured range for the g-factor matches that of an exciton comprising an electron and a hole residing in different valleys\cite{Koperski2019}. \textbf{(d)} PL spectra of trapped interlayer excitons collected on the same nanopillar as panels a and b at 0 T for $\sigma^{+}$-polarised excitation and co-polarised ($\sigma^{+}$) and cross-polarised ($\sigma^{-}$) detection, shown as black and red curves, respectively.}
\label{fig:3}
\end{figure*}

\begin{figure*}
\centerline{\includegraphics[width=180mm]{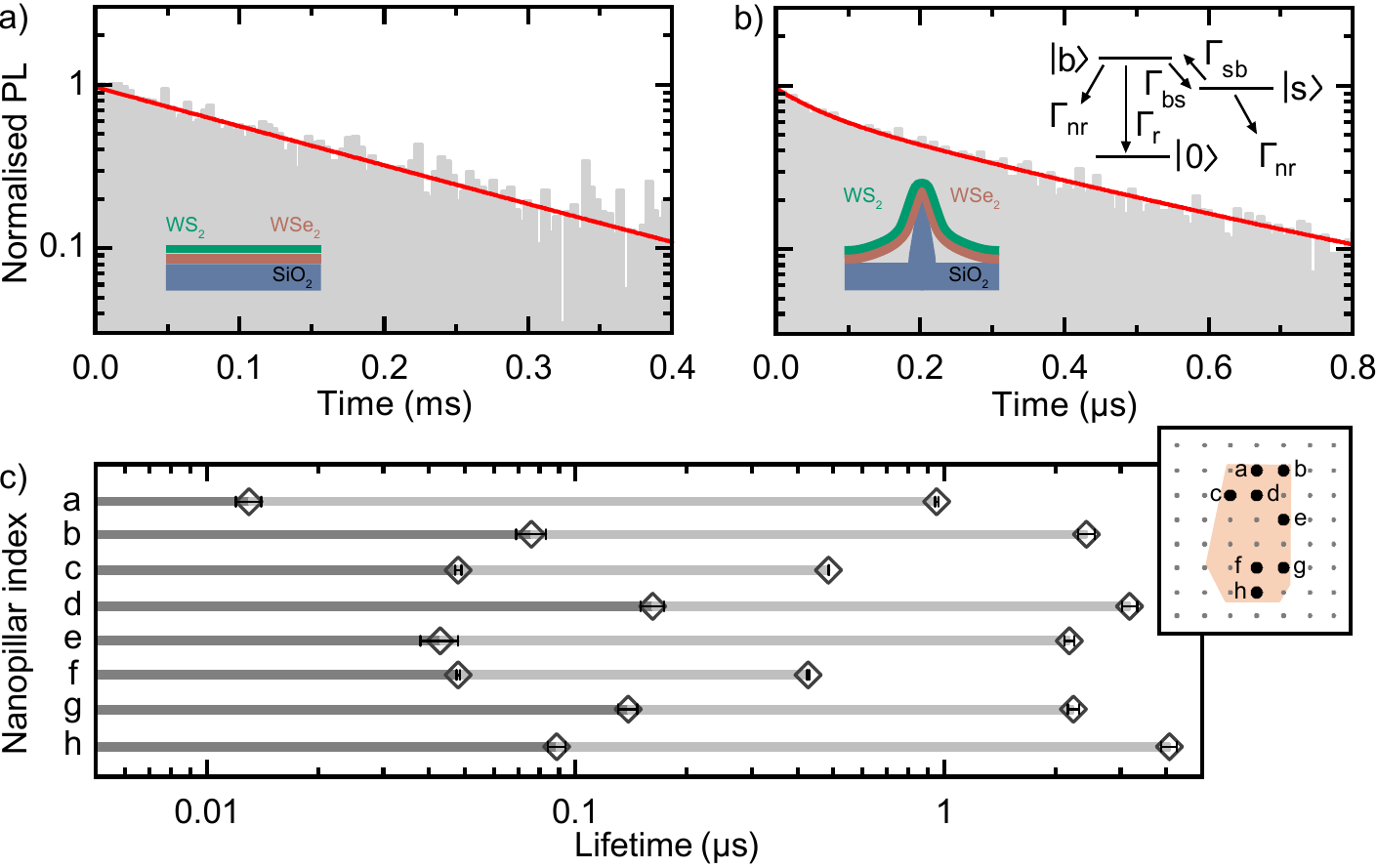}}
\caption{\textbf{Interlayer exciton lifetime.} \textbf{(a)} Lifetime measurement of interlayer exciton in the flat heterobilayer region at T=4 K. The data are fitted by a single exponential (solid red curve) with a lifetime $\tau=$180.6$\pm$3.5 $\mu$s. \textbf{(b)} Lifetime measurement of an example nanopillar-trapped interlayer exciton (1.39-eV peak from Fig. 3d). The solid red curve is a biexponential fit with two lifetimes $\tau_s=$59.4$\pm$0.9 ns and $\tau_l=$389.1$\pm$1.2 ns. The inset presents a three-level model with coupling between a bright state $\mid$b$\rangle$ and a shelving state $\mid$s$\rangle$, as discussed in Section 5 of the SI. \textbf{(c)} $\tau_s$ (dark grey bars) and $\tau_l$ (light grey bars) values extracted from biexponential fits to the lifetime measurements at the eight nanopillar locations (indexed a-h). The average value of $\tau_s$ ($\tau_l$) across all nanopillars is 80 ns (2 $\mu$s). The inset maps the physical location of the nanopillars on the heterobilayer region of our device.}
\label{fig:4}
\end{figure*}

\clearpage


\begin{center} 

{\large \textbf{Supplementary information for \\ Confinement of long-lived interlayer excitons in WS$_2$/WSe$_2$ heterostructures}}

\end{center}

\begin{center} 
Alejandro R.-P. Montblanch,$^{1,*}$ Dhiren M. Kara,$^{1,*}$ Ioannis Paradisanos,$^{2}$ Carola M. Purser,$^{1,2}$ Matthew S. G. Feuer,$^{1}$ Evgeny M. Alexeev,$^{2,1}$ Lucio Stefan,$^{1,3}$ Ying Qin,$^{4}$ Mark Blei,$^{4}$ Gang Wang,$^{2}$ Alisson R. Cadore,$^{2}$ Pawel Latawiec,$^{5}$ Marko Lon\v{c}car,$^{5}$ Sefaattin Tongay,$^{4}$ Andrea C. Ferrari,$^{2,\dag}$, and Mete Atat\"ure$^{1,\ddag}$
\end{center}

\begin{center}
\textit{\footnotesize{\hbox{$^{1}$Cavendish Laboratory, University of Cambridge, 19 J. J. Thomson Ave., Cambridge, CB3 0HE, UK}}}

\textit{\footnotesize{\hbox{$^{2}$Cambridge Graphene Centre, University of Cambridge, 9 J. J. Thomson Ave., Cambridge, CB3 0FA, UK}}}

\textit{\footnotesize{\hbox{$^{3}$The Faraday Institution, Quad One, Becquerel Avenue, Harwell Campus, Didcot, OX11 0RA, UK}}}

\textit{\footnotesize{\hbox{$^{4}$School for Engineering of Matter, Transport and Energy, Arizona State University, Tempe, AZ, 85287, USA}}}

\textit{\footnotesize{\hbox{$^{5}$School of Engineering and Applied Science, Harvard University, Cambridge, MA, 02138, USA}}}
\end{center}

\bigskip

\noindent \begin{small} \textbf{CONTENTS}

\noindent 1. Characterisation of WS$_{2}$/WSe$_{2}$ heterobilayer device \hfill 19

\noindent 2. Statistics of trapped interlayer excitons: power saturation and magnetic-field \\ dependence \hfill 20

\noindent 3. Power dependence of untrapped interlayer-exciton lifetime \hfill 21

\noindent 4. Spectrally selective lifetime measurements on a nanopillar \hfill 21

\noindent 5. Temperature dependence of trapped interlayer-exciton lifetime \hfill \hfill 22

\end{small}

\vfill
\par\noindent\rule{50pt}{0.4pt}

\noindent \footnotesize{$^{*}$These authors contributed equally.}

\noindent \footnotesize{$^{\dag}$acf26@cam.ac.uk}

\noindent \footnotesize{$^{\ddag}$ma424@cam.ac.uk}

\newpage

\begin{normalsize}

\subsection*{1. Characterisation of WS$_{2}$/WSe$_{2}$ heterobilayer device}


To characterise the sample structurally we perform room-temperature Raman measurements in a commercial Horiba LabRam Evolution system using a 514-nm laser at 60 $\mu$W power. Figure S1a shows the Raman spectrum from the WS$_{2}$/WSe$_{2}$ heterobilayer on a flat region. The Raman peaks at $\sim$358 and $\sim$419 cm$^{-1}$ correspond to the E$^{'}$ and A$^{'}_{1}$ modes of WS$_2$, respectively[1]. The separation between them $\sim$61 cm$^{-1}$ is consistent with it being monolayer[2]. The peak $\sim$251 cm$^{-1}$, with full-width at half-maximum $\sim$2 cm$^{-1}$, is assigned to the convoluted A$^{'}_{1}$+E$^{'}$ modes of WSe$_{2}$, degenerate in monolayer WSe$_2$[1,2], while the peak $\sim$262 cm$^{-1}$ belongs to the 2LA(M) mode of WSe$_2$[1].

Figure S1b plots the spatially resolved map of the spectrally integrated second-harmonic signal from monolayer and heterobilayer regions. This is measured with a picosecond-pulsed Ti:Sapphire laser ($\sim$3-ps pulse duration) at 76 MHz repetition rate and tuned to 982 nm at $\sim$1 mW CW-equivalent power. From the reduced intensity in the heterobilayer region compared to monolayer in Fig. S1b we infer a stacking angle between WSe$_{2}$ and WS$_{2}$ monolayers of $\sim$50$^{º}$, closer to 60$^{º}$ because the stacking of the individual constituents close to that angle yield partial destructive interference resulting from the combination of the second harmonic signal from each individual monolayer[3]. This is confirmed by additional polarisation-resolved second-harmonic generation measurements at 1300-nm pump wavelength. 

Figure S2a is the map of the amplitude retrace in the room-temperature atomic force microscope measurement (commercial head, model MFP-
3D from Asylum Research with silicon tips from Bruker) of an area of the device containing flat WSe$_{2}$ monolayer and WS$_{2}$/WSe$_{2}$ heterobilayer regions, as well as WS$_{2}$/WSe$_{2}$ heterobilayer on top of a nanopillar. Figure S2b shows the topography of the area enclosed by the white dashed square in Fig. S2a. Figure S2c plots the height profile obtained averaging the line cuts spanning a length of 350 nm centered around the blue arrow of Fig. S2b, and shows that the step size from the WSe$_{2}$ monolayer region to the WS$_{2}$/WSe$_{2}$ heterobilayer region coincides with the thickness of a monolayer, 0.6 nm[4].

\subsection*{2. Statistics of trapped interlayer excitons: power saturation and magnetic-field dependence}

Figure S3a is a set of power saturation curves of trapped interlayer excitons measured on different nanopillar locations at a temperature T=4 K, showing that the saturation behaviour of trapped interlayer excitons is reproducibly found across the device. We fit the data to the saturation function

\begin{equation}
A\times P^{n} /(P_{\rm sat} + P^{n})
\end{equation}

\noindent where \textit{P} is the excitation power, \textit{A} is the PL intensity at saturation, \textit{P}$_{\rm sat}$ is the saturation power, and \textit{n} is the exponent that describes the power law followed by \textit{P} at low excitation powers. We find a distribution of saturation powers with standard deviation $\pm$5 $\mu$W, indicative of a range of slightly dissimilar confining potentials. This is also supported by the observed variation in trapped exciton g-factors and lifetimes (Figs. 3c and 4c in the main text). We find an average value of \textit{n}=1.00$\pm$0.12. This linear dependence, reproduced across the device, suggests that the trapped interlayer excitons do not experience density-dependent decay mechanisms due to excitons around the nanopillar. This contrasts evidence of strong density dependence of the unbound interlayer excitons, which exhibits non-linear excitation-power dependence (Fig. 2b in the main text) as well as population-dependent lifetimes (Fig. S4). 

Figure S3b plots the Zeeman splitting of trapped interlayer exciton emission in two example nanopillar locations. The solid curves are linear fits to the data, from which we extract the g-factors of the trapped interlayer excitons according to[5]:

\begin{equation}
\label{eq:brightpop}
\Delta E=E_{\sigma_{-}}-E_{\sigma_{+}}=g\mu_{\rm B}B
\end{equation}

\noindent where $E_{\sigma_{+}}$ ($E_{\sigma_{-}}$) is the emission energy of a trapped interlayer exciton with $\sigma_{+}$ ($\sigma_{-}$) collection, $g$ is the trapped interlayer exciton g-factor, $\mu_{\rm B} = 58$ $\mu$eV T$^{-1}$ is Bohr's magneton and $B$ is the applied magnetic field. The g-factors obtained from the linear fits in Fig. S3b contribute to the Zeeman splitting distribution of similar g-factors with average value 13.2$\pm$1.1, as stated in the main text and represented in Fig. 3c by a blue-shaded area. The similarity in the g-factor distribution between both locations suggests that the magnetic states at each confining potential have the same microscopic origins, albeit not identical across a given nanopillar location.

\subsection*{3. Power dependence of untrapped interlayer-exciton lifetime}

We measure the lifetime of the untrapped interlayer excitons as a function of \textit{average} pump power, $\bar{P}$, at T=4 K. We adjust $\bar{P}$ by gating the pulsed laser (76-MHz repetition rate / 13-ns pulse separation) with varying gate windows from 50 ns up to 500 ns for a fixed histogram range of 1 ms. Figure S4 are lifetime measurements for three $\bar{P}$ in the linear pump power-PL intensity regime: $\bar{P}=2.5$ nW (black filled circles), $\bar{P}=7.5$ nW (red filled circles) and $\bar{P}=25$ nW (blue filled circles). The two lifetimes $\tau_1$ and $\tau_2$ quoted in Fig. S4 for each measurement are extracted from biexponential fits (solid curves). We observe that at larger pump powers a biexponential fit is required (see for example the red curve), while at lower pump powers a monoexponential fit suffices (as evidenced by the large fit error of the lifetime taken with the lowest $\bar{P}=2.5$ nW, black curve). This suggests a change in the dynamics of untrapped interlayer excitons with increasing pump powers. Additionally, in the linear excitation power regime, where the intensity scales linearly with $\bar{P}$, we observe a variation of both $\tau_1$ and $\tau_2$. This contrasts the behaviour of a non-interacting system, where the measured lifetimes would remain constant. Thus, this measurement supports the existence of density-dependent dynamics of untrapped interlayer excitons, as was already discussed for Fig. 2c of the main text.

\subsection*{4. Spectrally selective lifetime measurements on a nanopillar}

Figure S5 is a lifetime measurement on a nanopillar at T=4 K: the red (blue) dots corresponds to the red-shaded (blue-shaded) spectral region of the PL spectrum in the inset. The black dots are the lifetime measurement of the whole spectrum. We find that all measurements are best fit with a biexponential function (solid curves in Fig. S5), from which we extract $\tau_s=$24.2$\pm$4.0 ns and $\tau_l=$3.53$\pm$0.09 $\mu$s (red), $\tau_s=$104.6 $\pm$2.7ns and $\tau_l=$1.66$\pm$0.03 $\mu$s (blue), and $\tau_s=$159.9$\pm$8.8 ns and $\tau_l=$3.00$\pm$0.10 $\mu$s (black). Thus, the biexponential behaviour is maintained regardless of the selected spectral window, and is observed across all nanopillar locations.

While there is a slight variation, expected for distinct confining potentials across a nanopillar, lifetimes are within the same order of magnitude for different spectral regions and we do not capture any correlation between spectrum and lifetime. The biexponential is observed down to the single-peak level, as presented in the lifetime measurement in Fig. 4b of the spectrally isolated peak at 1.39 eV in Fig. 3d of the main text.

The close similarity between the lifetime of the spectrally selected regions and the lifetime extracted for the integrated PL spectrum, combined with the similarity in intensities among all spectral peaks for a given nanopillar, prevents the distinction of each individual biexponential decay for each single peak in a lifetime measurement of the integrated PL spectrum on a nanopillar. The measured lifetimes are thus representative of the average lifetimes for trapped interlayer excitons at each nanopillar location, and justifies our reporting of the average lifetime from each nanopillar location in Fig. 4c as extracted by collecting the PL from the entire, integrated PL spectrum.

\subsection*{5. Temperature dependence of trapped interlayer-exciton lifetime}

To explain the biexponential decay of the trapped interlayer excitons we start with a three-level model, depicted in Fig. S6a, consisting of a bright state $\mid$b$\rangle$, an optically inactive shelving state $\mid$s$\rangle$ and a ground state $\mid$0$\rangle$. Similar models have been used earlier in other systems with biexponential decay characteristics, such as self-assembled quantum dots[6] and carbon nanotubes[7].

The rate equations governing the populations $\rho_{b}$ and $\rho_{s}$ of the bright and shelving states in such a system are:

\begin{equation}
\label{eq:brightpop}
\frac{\partial \rho_{b}}{\partial t} = -(\Gamma_{r}+\Gamma_{nr}+\Gamma_{bs})\rho_{b}+(\Gamma_{sb})\rho_{s}
\end{equation}

\begin{equation}
\label{eq:shelvpop}
\frac{\partial \rho_{s}}{\partial t} = -(\Gamma_{nr}+\Gamma_{sb})\rho_{s}+(\Gamma_{bs})\rho_{b}
\end{equation}

\noindent where the radiative rate for the bright state is $\Gamma_{r}$, the coupling rates between the bright and the shelving states are $\Gamma_{bs}$ (bright to shelving) and $\Gamma_{sb}$ (shelving to bright), and the nonradiative decay rate for the bright and shelving states (assumed to be equal for simplicity) is $\Gamma_{nr}$.

Solving the system of linear differential Eqs. (3,4) yields a biexponential population decay in time for $\rho_{b}(t)$ of the form A$_s$e$^{-t/\tau_{s}}$+A$_l$e$^{-t/\tau_{l}}$ as written in the main text, with time constants given by

\begin{equation}
\label{eq:timeconst1}
\frac{\Gamma_{r}+\Gamma_{bs}+\Gamma_{sb}}{2} + \Gamma_{nr} \pm \sqrt{ \left( \frac{\Gamma_{r}+\Gamma_{bs}-\Gamma_{sb}}{2} \right)^{2} + \left( \frac{\Gamma_{bs}+\Gamma_{sb}}{2}\right)^{2} - \left( \frac{\Gamma_{bs}-\Gamma_{sb}}{2} \right)^{2} }.
\end{equation}

We identify the fast decay rate $\Gamma_{s}=\frac{1}{\tau_{s}}$ with the + sign of Eq. (5) and the slow decay rate $\Gamma_{l}=\frac{1}{\tau_{l}}$ with the - sign of Eq. (5).

We perform temperature-dependent lifetime measurements to elucidate the dominating rate and gain insight on the decay mechanisms. Figure S6b is an example lifetime measurements taken on a nanopillar at two different temperatures: 4 K (black filled circles) and 70 K (red filled circles). We extract, for an example nanopillar location, the two exponents of the biexponential fit at different temperatures, and present them as rates 1/$\tau_{s}$ (fast decay rate) and 1/$\tau_{l}$ (slow decay rate) in Fig. S6b as orange and black dots, respectively. We observe a strong dependence of both 1/$\tau_{s}$ and 1/$\tau_{l}$ with temperature, suggesting that $\Gamma_{nr}$, $\Gamma_{bs}$ and $\Gamma_{sb}$ (Fig. 4 in the main text) are thermally activated: a common mechanism for nonradiative processes, as well as coupling between a bright and a shelving state, is through phonon mediation. To model this coupling we assume a multi-phonon process involving a continuum phonon density of states[8] and set the temperature dependences for $\Gamma_{nr}$, $\Gamma_{bs}$ and $\Gamma_{sb}$ to be, respectively:

\begin{equation}
\label{eq:gammanrtemp}
\Gamma_{nr}(T) = \gamma_{nr} \left( 1+\alpha T^{m} \right) 
\end{equation}

\begin{equation}
\label{eq:gammabstemp}
\Gamma_{bs}(T) = \gamma_{bs} \left( 1+\beta T^{m} \right) 
\end{equation}

\begin{equation}
\label{eq:gammabstemp}
\Gamma_{sb}(T) = \gamma_{bs} \beta T^{m} 
\end{equation}

\noindent where $T$ is temperature, $\gamma_{nr}$ is the spontaneous non-radiative rate at 0 K, $\alpha$ is the pre-factor of $T^{m}$ for $\Gamma_{nr}$, $\gamma_{bs}$ is the rate of population transfer from the bright to the shelved state at 0 K, $\beta$ is the pre-factor of $T^{m}$ for both $\Gamma_{bs}$ and $\Gamma_{sb}$, and $m$ describes the power law scaling of $\Gamma_{nr,bs,sb}$ with $T$. $m$ depends on the exact phononic coupling mechanism[8], and we assume for simplicity to be the same for all $\Gamma_{nr,bs,sb}$. The solid curves in Fig. S6b are fits to the data obtained by substituting Eqs. (6-8) into Eq. (5), where $\Gamma_{r}$ is left as a constant in $T$ because it physically depends only on the optical dipole transition strength. We obtain $m\approx$3, a value that has been obtained before in other quantum emitters[8] and confirms the strong temperature dependence of $\Gamma_{nr}$, $\Gamma_{bs}$ and $\Gamma_{sb}$. The remaining fit values are: $\Gamma_{r}=20.1\pm7.4$ MHz, $\gamma_{nr}=2.77\pm0.94$ MHz, $\alpha=\left( 6.0\pm3.7 \right)\times 10^{-6}$ K$^{-3}$, $\gamma_{bs}=3.8\pm6.3$ MHz and $\beta= \left( 2.1\pm3.8 \right)\times10^{-5}$ K$^{-3}$.

With these values, $\Gamma_{sb}$ has a meaningful contribution starting at $\sim$20 K, which corresponds to an energy of $\sim$2 meV, and $\Gamma_{sb}$ becomes comparable in magnitude to $\Gamma_{bs}$ at $\sim$40 K. Our model reveals that at low temperatures (below 20 K), when $\Gamma_{sb}$ is small, Eq. (5) can be approximated as:

\begin{equation}
\label{eq:gammabstemp}
\Gamma_{l}\approx \Gamma_{nr}
\end{equation}

\begin{equation}
\label{eq:gammabstemp}
\Gamma_{s}\approx \Gamma_{r}+\Gamma_{bs}+\Gamma_{nr}
\end{equation}

In this case, $1/\tau_{l}=\Gamma_{l}$ in the biexponential fit of a lifetime measurement becomes a direct measure of $\Gamma_{nr}$ in the limit of low temperature. Because $\tau_{l}$ is more than one order of magnitude larger than $\tau_{s}$ in our measurements (Fig. 4c in the main text),  $\Gamma_{nr}$ is small compared at least to the sum $\Gamma_{r}+\Gamma_{bs}$ indicating that, at low temperatures, the dynamics of trapped interlayer excitons is not limited by nonradiative processes. In particular, applying the fit parameters in Eqs. (6-8) we obtain, at 4 K, $\Gamma_{r}$=20.1 MHz, $\Gamma_{nr}$=2.8 MHz and $\Gamma_{bs}$=3.8 MHz: $\Gamma_{r}$ is roughly one order of magnitude larger than $\Gamma_{bs}$ and $\Gamma_{nr}$, so that $1/\tau_{s}=\Gamma_{s}$ is an approximate measure of $\Gamma_{r}$ in the limit of low temperature. We note that it is the long-lived shelving state that enables this relation: a model with only one excited state would yield a single exponential and not two. Additionally, the measured lifetimes up to 4 $\mu$s (Fig. 4c in the main text) indeed indicate that a strong fraction of the trapped interlayer excitons will live for that long, regardless of the mechanism that enables this. At 4 K we find, from the values in Fig. 4c in the main text, that the average radiative rate across nanopillars is $\Gamma_{r}\approx$13 MHz, and the average non-radiative rate is $\Gamma_{nr}\approx$500 kHz.

At temperatures beyond $\sim$40 K, when $\Gamma_{bs}\approx\Gamma_{sb}$, the model becomes:

\begin{equation}
\label{eq:gammabstemp}
\Gamma_{s}\approx \frac{\Gamma_{r}}{2}+ \Gamma_{bs} + \Gamma_{nr} + \sqrt{ \left( \frac{\Gamma_{r}}{2} \right)^{2} + \left( \Gamma_{bs} \right)^{2} }
\end{equation}

\begin{equation}
\label{eq:gammabstemp}
\Gamma_{l}\approx \frac{\Gamma_{r}}{2}+ \Gamma_{bs} + \Gamma_{nr} - \sqrt{ \left( \frac{\Gamma_{r}}{2} \right)^{2} + \left( \Gamma_{bs} \right)^{2} }
\end{equation}

With our fit parameters we find, at 70 K, $\Gamma_{r}$=20.1 MHz, $\Gamma_{nr}$=8.5 MHz, $\Gamma_{bs}$=31.2 MHz and $\Gamma_{sb}$=27.4 MHz, so that nonradiative processes and population transfer to (and from) the shelving state are at this temperature more relevant. Our model predicts the decrease in quantum efficiency with increasing temperature. Indeed, the thermally induced modification in the dynamics is evidenced in Fig. S6d: it plots the PL intensity of trapped interlayer excitons as a function of temperature, showing that as nonradiative processes are thermally activated, the PL intensity decreases.

Finally, we note that an exponential temperature dependence of the form $\exp{\left( -\frac{E_{a}}{k_{B}T }\right)}$ (with $E_{a}$ the activation energy, $k_{B}$ the Boltzmann constant and $T$ temperature), used in other systems in the context of the Mott-Seitz model[9], also fits our data reasonably with similar parameter values. Thus, while the three-level model with temperature-dependent rates captures the trapped interlayer exciton dynamics successfully, a thorough theoretical study and further experimental analysis is necessary to understand the full microscopic nature of the system.

\noindent \subsection*{References}

\noindent [1] Terrones, H. \textit{et al.} New first order Raman-active modes in few layered transition metal dichalcogenides. \textit{Sci. Rep.} \textbf{4}, 4215 (2014).

\noindent [2] Zhao, W. \textit{et al.} Lattice dynamics in mono- and few-layer sheets of WS$_2$ and WSe$_2$. \textit{Nanoscale} \textbf{5}, 9677 (2013).

\noindent [3] Hsu, W.-T. \textit{et al.} Second Harmonic Generation from Artificially Stacked Transition Metal Dichalcogenide Twisted Bilayers. \textit{ACS Nano} \textbf{8}, 2951-2958 (2014).

\noindent [4] Rosenberger, M \textit{et al.} Nano-"Squeegee" for the Creation of Clean 2D Material Interfaces. \textit{ACS Appl. Mater. Interfaces} \textbf{10}, 10379 (2018).

\noindent [5] Aivazian, G. \textit{et al.} Magnetic control of valley pseudospin in monolayer WSe$_{2}$. \textit{Nat. Phys.} \textbf{11}, 148 (2015).

\noindent [6] Tighineanu, P. \textit{et al.} Decay dynamics and exciton localization in large GaAs quantum dots grown by droplet epitaxy. \textit{Phys. Rev. B} \textbf{88}, 155320 (2013).

\noindent [7] Berciaud, S. \textit{et al.} Luminescence Decay and the Absorption Cross Section of Individual Single-Walled Carbon Nanotubes. \textit{Phys. Rev. Lett.} \textbf{101}, 077402 (2008).

\noindent [8] Jahnke, K. D. \textit{et al.} Electron–phonon processes of the silicon-vacancy centre in diamond. \textit{New J. Phys.} \textbf{17}, 043011 (2015).

\noindent [9] Toyli, D. M. \textit{et al.} Measurement and Control of Single Nitrogen-Vacancy Center Spins above 600 K. \textit{Phys. Rev. X} \textbf{2}, 031001 (2012).

\newpage

\captionsetup[figure]{labelformat=empty}

\begin{figure*}
\centerline{\includegraphics[width=170mm]{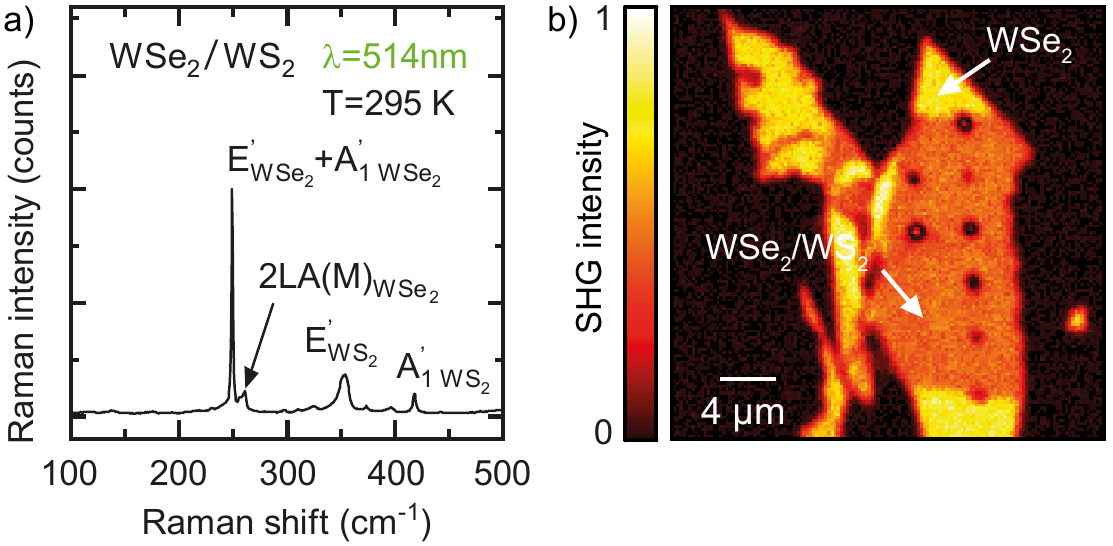}}
\caption*{FIG. S1: \textbf{Raman and second harmonic signal of the WS$_2$/WSe$_2$ device.} \textbf{(a)} Room-temperature Raman spectrum of the heterobilayer device, confirming the monolayer nature of each constituent layer. \textbf{(b)} Spatially resolved intensity of the second-harmonic signal of the device with pump laser at 982 nm.}
\label{fig:S1}
\end{figure*}

\begin{figure*}
\centerline{\includegraphics[width=150mm]{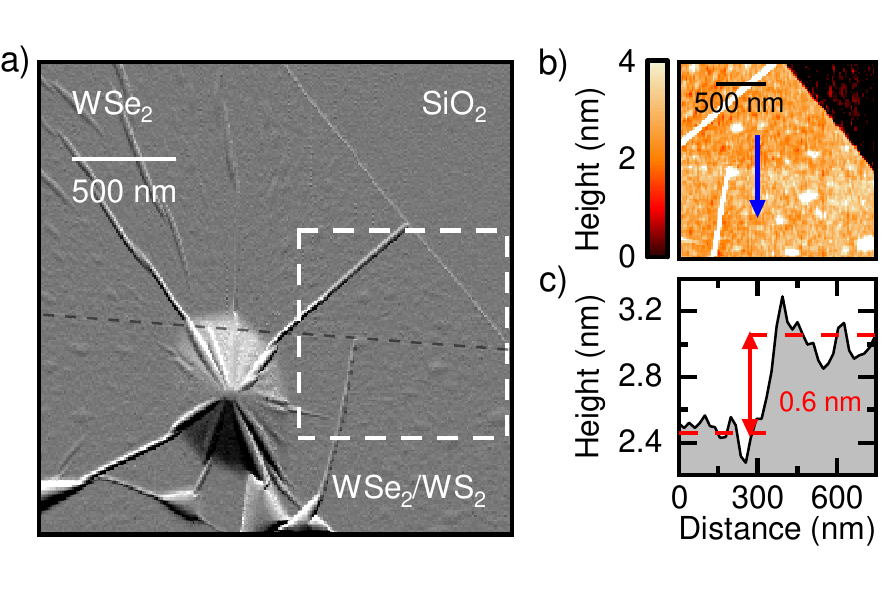}}
\caption*{FIG. S2: \textbf{Atomic force microscope image of the WS$_2$/WSe$_2$ device.} \textbf{(a)} Amplitude retrace of a device area with SiO$_{2}$, WSe$_{2}$ monolayer and WS$_2$/WSe$_2$ heterobilayer regions, where the black dashed line marks the boundary between the last two regions. \textbf{(b)} Topography of the area enclosed by the white dashed square in a. \textbf{(c)} Height profile obtained as the average height across the line cuts spanning a length of 350 nm centered around the blue arrow in b.}
\label{fig:AFM}
\end{figure*}

\begin{figure*}
\centering
\centerline{\includegraphics[width=170mm]{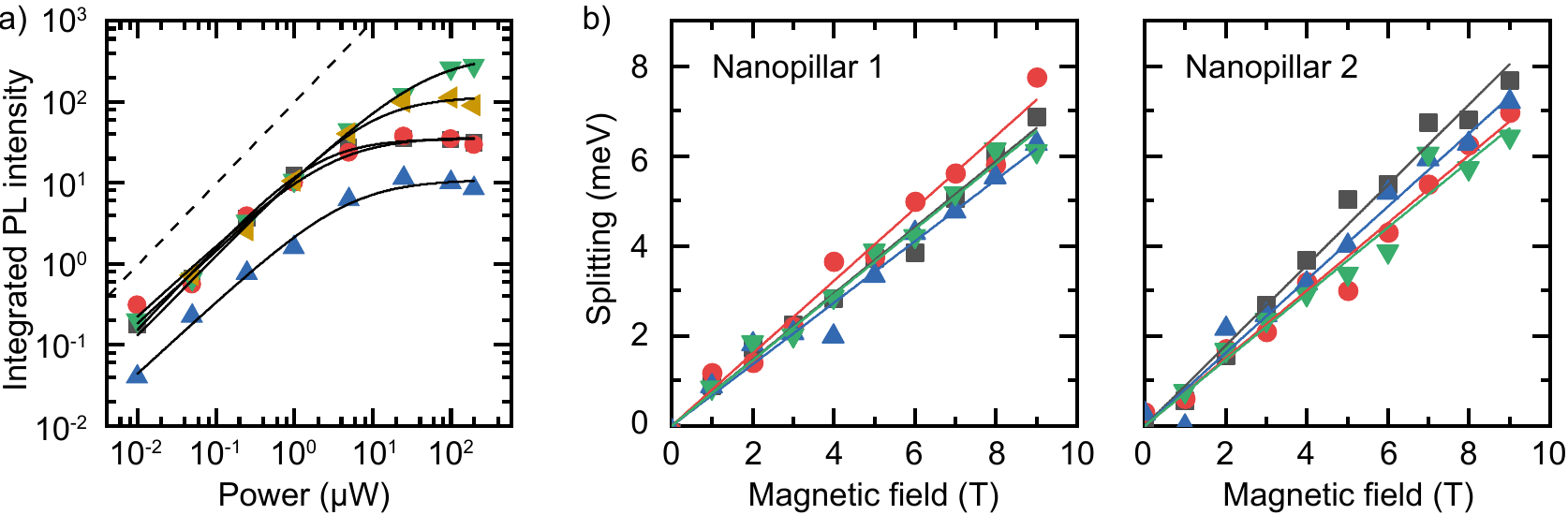}}
\caption*{FIG. S3: \textbf{Saturation curves and Zeeman splittings of trapped interlayer excitons across different pillar locations.} \textbf{(a)} Saturation curves under CW excitation of trapped interlayer excitons across different nanopillar locations. The solid lines are fits using Eq. (1), and the dotted line corresponds to a linear excitation power dependence. All fits shown here present an initial linear dependence with the excitation power \textit{P} (generally up to the order of a few $\mu$W), and then saturate. \textbf{(b)} Zeeman splitting of trapped different interlayer exciton emission in two separate nanopillar locations. Curves are linear fits to the Zeeman splittings, each one corresponding to the various trapped interlayer excitons within a given nanopillar location.}
\label{fig:SatCurves}
\end{figure*}

\begin{figure*}
\centerline{\includegraphics[width=90mm]{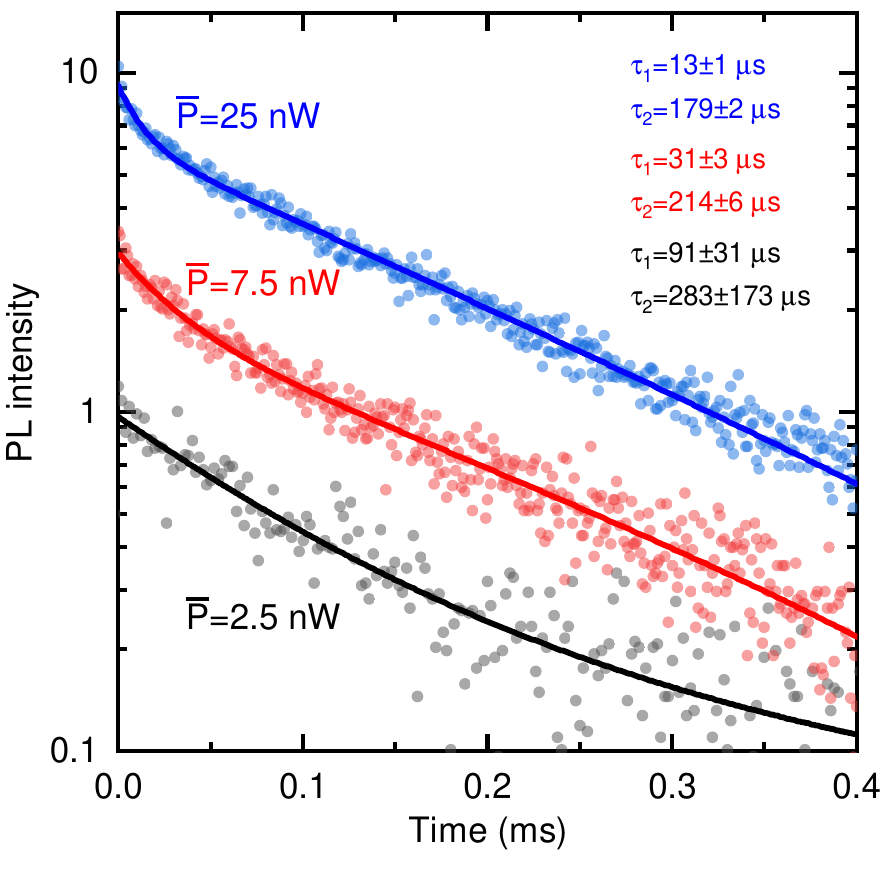}}
\caption*{FIG. S4: \textbf{Lifetime of untrapped interlayer exciton for different average pump powers $\bar{P}$}. Filled circles are data points and solid lines are biexponential fits. The quoted lifetimes $\tau_1$ and $\tau_2$ are extracted from the fits of the three measurements shown.}
\label{fig:S5}
\end{figure*}

\begin{figure*}
\centerline{\includegraphics[width=90mm]{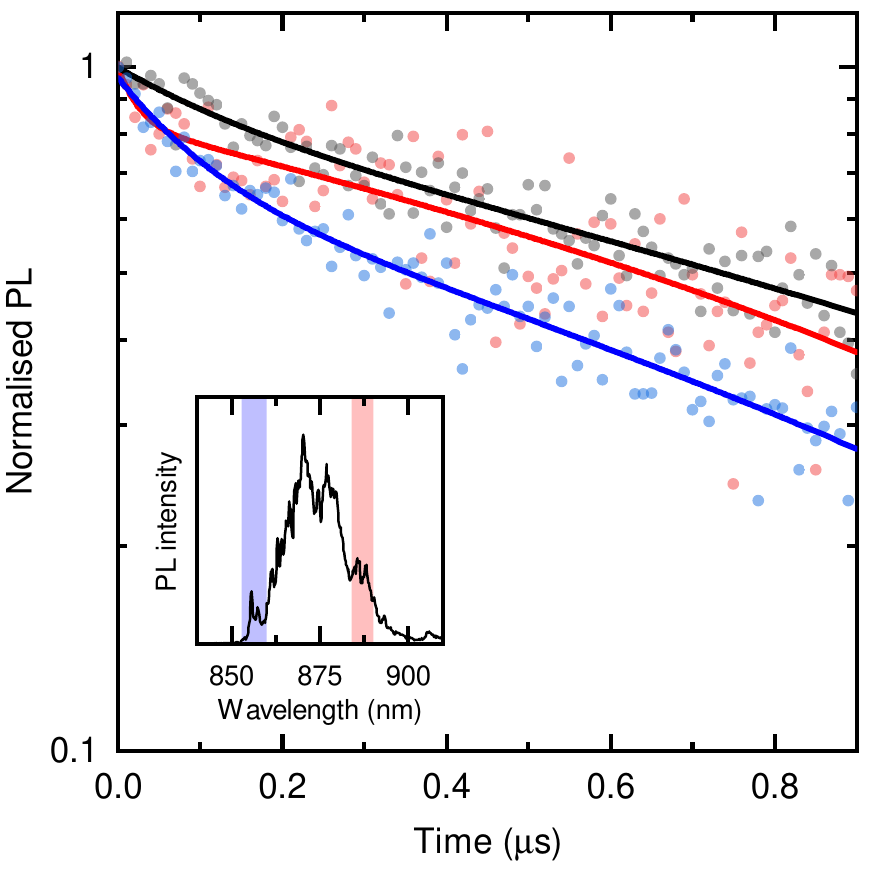}}
\caption*{FIG. S5: \textbf{Lifetime measurement of three spectral regions of the PL signal on a nanopillar}. Dots are data and solid lines are biexponential fits. Lifetimes in read and blue correspond to the spectral regions shaded in red and blue, respectively, in the PL spectrum in the inset. The lifetime of the whole PL spectrum in the inset is in black.}
\label{fig:S4}
\end{figure*}

\begin{figure*}
\centerline{\includegraphics[width=170mm]{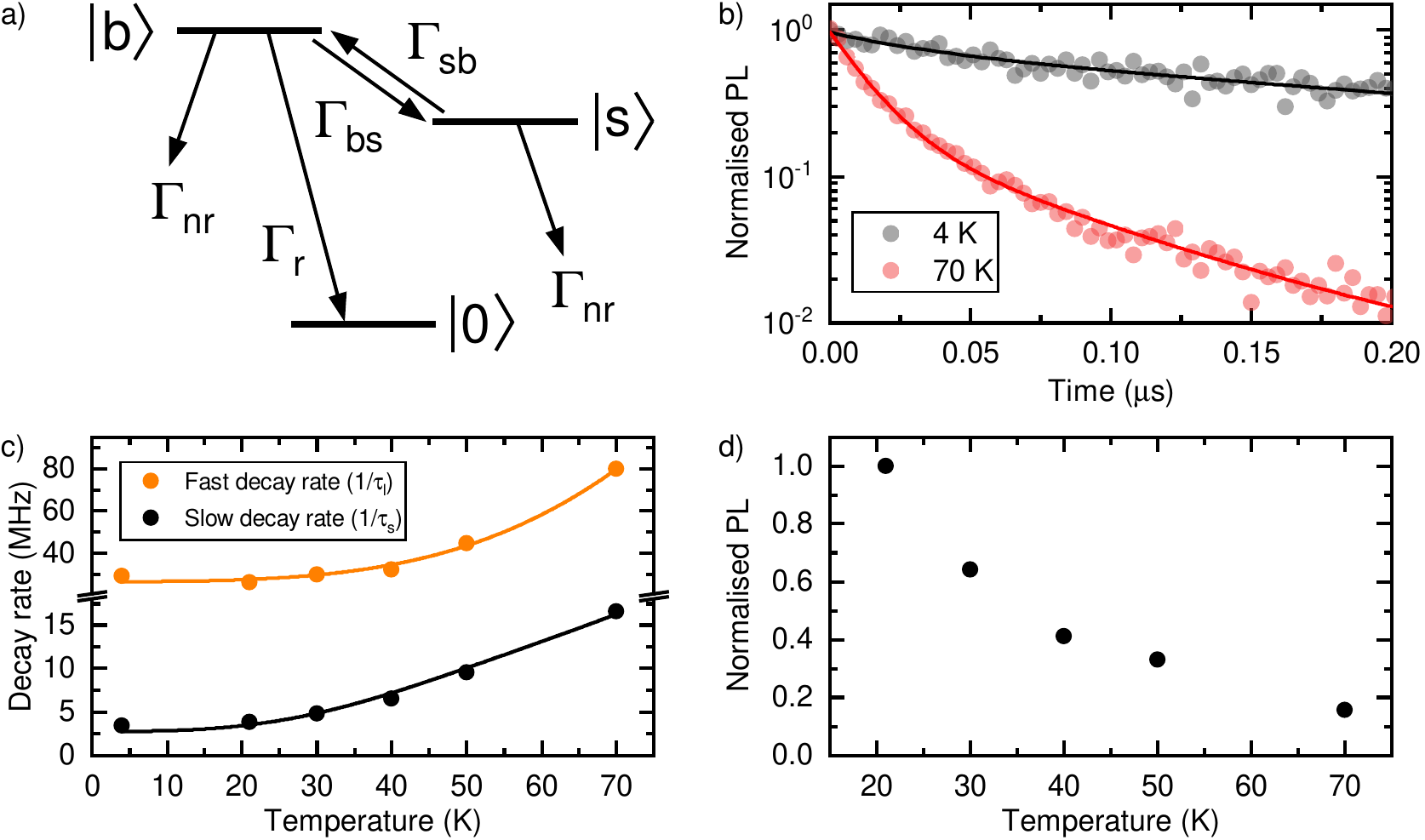}}
\caption*{FIG. S6: \textbf{Three-level model and lifetimes of trapped interlayer excitons as a function of temperature}. \textbf{(a)} Schematic representation of the three-level model used to explain the biexponential behaviour of trapped interlayer excitons. The meaning of each rate is specified in the text. \textbf{(b)} Lifetime measurement of an example nanopillar-trapped interlayer exciton at two different temperatures: 4 K (black dots) and 70 K (red dots). The black and red curves are biexponential fits to the data at 4 K and 70 K, respectively. \textbf{(c)} Decay rates as a function of temperature obtained from biexponential fits to lifetime measurements on an example nanopillar-trapped interlayer exciton. The fast (slow) decay rate 1/$\tau_{s}$ (1/$\tau_{l}$) is represented by orange (black) dots. The solid curves are fits obtained using our model described in the text. \textbf{(d)} PL intensity on an example nanopillar-trapped interlayer exciton as a function of temperature.}
\label{fig:S6}
\end{figure*}

\end{normalsize}

\end{document}